\documentclass[twocolumn,prl,superscriptaddress,floatfix,showpacs]{revtex4-1}
\usepackage[utf8]{inputenc}
\usepackage{amsmath}
\usepackage{amssymb}
\usepackage{graphicx}

\makeatletter
\usepackage{graphics}
\usepackage{epsfig}
\usepackage{color}

\newcommand{\be}{\begin{equation}} \newcommand{\ee}{\end{equation}}
\newcommand{\bea}{\begin{eqnarray}} \newcommand{\eea}{\end{eqnarray}}

\def\alain#1{\textcolor{black}{#1}}		
\def\alnew#1{\textcolor{black}{#1}}		
\def\alnewnew#1{\textcolor{black}{#1}}		
\def\matteo#1{\textcolor{black}{#1}}		

\makeatother

\begin{document}

\title{Synchronous population activity for optimal heterogeneity of inhibitory neurons  in spiking neural networks. }

\title{\alain{Optimal responsiveness and new activity states emerging from} the heterogeneity of inhibitory neurons.}

\title{Optimal responsiveness and collective oscillations emerging from the heterogeneity of inhibitory neurons.}

\author{Matteo di Volo} 
\affiliation{Laboratoire de Physique Th\'eorique et Mod\'elisation, Universit\'e de Cergy-Pontoise, CNRS, UMR 8089,
95302 Cergy-Pontoise cedex, France}
\author{Alain Destexhe} 
\affiliation{Paris-Saclay University, Institute of Neuroscience, CNRS, Gif sur Yvette, France}

\date{\today}

\begin{abstract}

The brain is characterized by a strong heterogeneity of inhibitory neurons.  We report that spiking neural networks display a resonance to the heterogeneity of inhibitory neurons, with optimal input/output responsiveness occurring for levels of heterogeneity similar to that found experimentally in cerebral cortex. A heterogeneous mean-field model predicts such optimal responsiveness. Moreover, we show that new dynamical regimes emerge from heterogeneity that were not present in the equivalent homogeneous system, such as sparsely synchronous collective oscillations.


\end{abstract}

\maketitle


Studying the collective behavior of large numbers of units interacting non-linearly is a classical theme in physical sciences.  In biology, such studies are complicated by the fact that the units are usually non identical, but rather display considerable heterogeneity.  This particularly apparent in cerebral cortex, where neuronal size and properties are highly heterogeneous \cite{tsodyks1993pattern,roxin2011distribution,roxin2011role,denker2004breaking,neltner2000synchrony,golomb1993dynamics,zerlaut2016heterogeneous,tseng1993heterogeneity,pospischil2008minimal,sharpee2014toward,landau2016impact}. Neuronal heterogeneity is particularly high for inhibitory neurons, for which many cell classes were observed \cite{Gupta2000,Defelipe2013,Jiang2015,Mihaljevic2019}. In networks of oscillators, as in neuronal networks, such heterogeneity across units (bare frequencies or neurons' excitabilities) induces typically desynchronization at a population scale \cite{chow1998phase,white1998synchronization,traub2005single,wang1996gamma,tiesinga2000robust,devalle2018dynamics,luccioli2019neural}. As one would expect, the more the neurons are different the less they are able to synchronize and to correlate their reciprocal activity.
Nevertheless, despite this heterogeneity, cortical populations are able to respond coherently to external stimuli and generate collective oscillations \cite{gray1994synchronous,llinas1993coherent}.

In this paper, we examined networks of excitatory and inhibitory neurons, with an emphasis on inhibitory heterogeneity to understand its possible impact at the population level.  We used networks
of $N$ neurons, 80\% of excitatory ($N_E=0.8N$) and 20\% of inhibitory ($N_I=0.2N$) neurons. The membrane potential $V_i$ of each neuron evolves \alnew{according to} the adaptive exponential integrate and fire model \cite{brette2005adaptive}:
\begin{align}
&C_{m}\dot V_i = g_L(E_L^i-V_i)+g_Le^{\frac{V-v_{th}}{\Delta}}+I_{ext}+I_{s}^i-w_i  \\
&\tau_w\dot w_i =-w_i + b\sum_{\{ t_i^ {sp}\}} \delta(t- t_i^ {sp}) + a(V_i - E_L^i),
\label{network}
\end{align}
where $C_m=200$pF is the membrane capacitance, $g_L = 15$nS the leakage conductance, $v_{th}=50$mV the effective threshold and $\Delta$=0.5mV defines the action potential rise. The adaptation current $w_i$ increases of an amount $b=40$nS at each spike emitted by neuron $i$ at times $\{ t_i^ {sp}\}$ and has an exponential decay with time scale $\tau_w=500$ms. The parameter $a$ indicates subthreshold adaptation, that we will consider $a=0$nS if not otherwise stated. The current $I_{ext}$ an external current and $I_{s}$ the synaptic current from other neurons. We consider a random graph where each couple of neurons is connected with probability $p=0.05$. By calling $\{ t_j^ {sp}\}$ the ensemble of  spiking times of neuron $j$ the synaptic current received by neuron $i$ evolves as
\begin{align}
&I_{s}^i= g_i^E(V_i-E_E)+g_i^I(V_i-E_I)\\
&\tau_{s}\dot g_i^{E,I}=g_i^{E,I}+\frac{Q_{E,I}}{N}\sum_{\{ t_j^ {sp}\}\in(E,I)}\delta(t- t_j^ {sp}),
\label{network_1}
\end{align}
where $E_{E,I}$ is the reversal for excitatory ($E_{E}=0$mV) and inhibitory synapses ($E_{I}=-80$mV), $\tau_{s}=5$ms the synaptic decay time and $Q_{E,I}$ the interaction strength of excitatory ($Q_{E}=10\mu$S) and inhibitory ($Q_{I}=50\mu$S) synapses. The sum runs over the presynaptic excitatory or inhibitory neurons. The current $I_{ext}$ is an external input received by all neurons, independent and identically distributed, modeled as an additive excitatory Poissonian spike train at a frequency $\nu_{ext}$. In absence of stimuli  $\nu_{ext}$ is constant in time and we consider $\nu_{ext}=2.5$Hz to  keep the network active.  In order to study the response to external stimuli a time varying  $\nu_{ext}(t)$ will be considered (see Figure \alnew{captions} for details).

To model heterogenetity, we consider\alain{ed} a Gaussian distribution of the leakage reversal $E_L^i$ of inhibitory (excitatory) population $ \mathcal{N}(E_L^{E,I},\,\sigma_{E,I}^{2})$. The rescaled standard deviation $\sigma_{E,I}/E_L^{E,I}$ will be the main parameters to quantify heterogeneity.  The synchronization of the network and its response to external stimuli are quantified by measuring the total amount of evoked spikes $R$ (responsiveness) and  the input/output correlation between the stimuli and excitatory neurons firing rate $r^E$, $C=\langle r^E\nu_{ext} \rangle-\langle r^E \rangle\langle \nu_{ext} \rangle$, where $\langle \cdot \rangle$ denotes a time average.
 

\begin{figure}
\begin{centering}
\includegraphics[width=0.53\textwidth,clip=true]{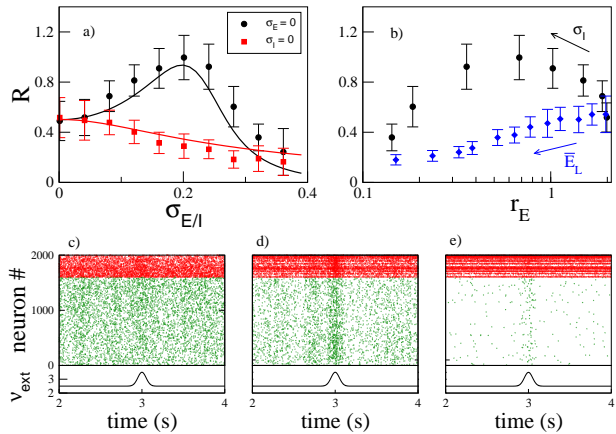}
\end{centering}
\vspace{-20pt}
\caption{\label{fig:1} Panel a) The evoked response $R$ to an external stimuli is reported in function of the heterogeneity of inhibitory (excitatory) neurons $\sigma_I$ ($\sigma_E)$, black (red) dots (squares). Error bars are estimated as the standard deviation over 20 different realisations. Continuous lines report the prediction based on the mean field model (see main text). Panel b) Data in a), black dots, are reported in function of the spontaneous excitatory firing rate $r^E$ before the input arrival (average over $10$s). Blue diamonds report $R$ as measured by varying the average leakage current of inhibitory neurons $\bar E_L$ for $\sigma_I=\sigma_E=0$. Lower panels show the the raster plot, i.e. spiking times of excitatory (inhibitory) neurons marked with green (red) dots, for c) $\sigma_I=0$, d) $\sigma_I=0.2$ and e) $\sigma_I=0.4$ . Lower inset shows the time course of $\nu_{ext}$. In this simulation we consider $N=10000$ neurons.}
\end{figure}

A first effect of inhibitory heterogeneity is to increase population responsiveness, as shown in Fig.~\ref{fig:1}.  In this simulation, the homogeneous network shows a
spontaneous asynchronous activity where neurons fire irregularly, a dynamical regime due to the balance between excitation and inhibition typically observed in the cortex of awake animals, called asynchronous irregular \cite{monteforte2010dynamical,van1996chaos,destexhe2003high}. The presence of external excitatory stimuli of short duration (see caption of Fig.~1 for details) produces a transient synchronous event (population burst). The total amount of additional spikes evoked by the stimuli is indicated by $R$. By increasing the amount of heterogeneity in inhibitory neurons $\sigma_I$ we observe a \alnew{clear} increase of the evoked activity $R$ (see Fig.~1a). Indeed, the same input induces a bell-shaped response $R$ in function of the heterogeneity  $\sigma_I$.  As can be seen \alnew{from Fig.~1c,d,e}, there is an optimal heterogeneity level ($\sigma_I\sim 0.2$), where the stimuli provokes a strong synchronous population response where almost the whole network activates. Eventually, when $\sigma_I$ is too large the response is very weak. As can be noticed from \alnew{the} bottom panels of Fig.~1, increasing heterogeneity in inhibitory neurons decreases excitatory neurons spontaneous activity. This is due to the presence of a fraction of inhibitory neurons with high excitability that inhibits the excitatory population. In Fig.~1b we report $R$, as in Fig.~1a, \alnew{as a} function of the excitatory spontaneous activity pre--stimulus $r_E$. We can observe that the responsiveness is maximum, in correspondence of $\sigma_I\sim 0.2$, for a relatively low value of excitatory spontaneous activity (around $r_E\sim 0.7$Hz). Interestingly, if we consider a homogeneous network, i.e. $\sigma_I=0$, and decrease spontaneous excitatory activity by increasing the average excitability of inhibitory neurons  we do not observe an increase of responsiveness $R$ (see blue dots in Fig.~1b). This result shows that the increase in the size of the synchronous response is due to the presence of heterogeneity and cannot be replaced by a modification of \alnew{the} average excitability in the corresponding homogeneous network.

In contrast, increasing \alnew{the} heterogeneity \alnew{of} excitatory neurons $\sigma_E$ has the opposite effect, i.e. a constant decrease in network response (red squares in Fig.~1a). At first sight, this seems contradictory with previous reports where it was shown that the heterogeneity of excitatory neurons' intrinsic excitability can induce  higher population responsiveness \cite{mejias2012optimal,gollo2016diversity}. However, in all these studies, the homogeneous system was set in an  excitable state characterised by no spontaneous activity in the absence of stimulus. Such an excitable regime is qualitatively different from the one we consider\alain{ed} in Fig.~1, where all neurons are characterised by spontaneous and irregular ongoing activity, \alnew{like in cerebral cortex} when excitation and inhibition balance each other.  Could this be the origin of the difference we observed?

To verify this scenario, we decreased the constant excitatory drive $\nu_{ext}$ that all neurons receive, in such a way that the homogeneous network is silent (excitable) in \alnew{the} absence of external stimuli. We then considered the correlation $C$ between the external input (that activates the network) and the network output $r^E$. We report in Fig.~2 the correlation $C$ in function of both $\sigma_E$ and $\sigma_I$. We observ\alain{e} that increasing $\sigma_E$ we recover an increase of the correlation $C$ that decreases for high $\sigma_E$, just like in previous studies \cite{mejias2012optimal,gollo2016diversity}. Indeed, increasing heterogeneity \alnew{among} excitatory neurons permits the presence of active neurons in response to the stimuli (see the white line in Fig.~2 separating silent and active spontaneous activity pre--stimulus). We report here, on top of an optimal response in the direction of $\sigma_E$, an optimal responsiveness also in the direction of $\sigma_I$. Indeed, once the network is active (i.e. high $\sigma_E$) the role of $\sigma_I$ is the same of that we reported in Fig.~1a, in which case the network was active due to a higher constant external drive. Eventually, when heterogeneity is too large, neurons are too diverse and cannot correlate in order to yield a coherent synchronous population response to the stimuli. \alain{We have  verified this scenario to be robust across the specific choice of the stimuli (amplitude, frequency etc..). }

This result shows for the first time an optimal amount of heterogeneity in both excitatory and inhibitory neurons for a coherent population response to external stimuli.  \alain{But is this optimal heterogeneity comparable to what is found experimentally?  To answer this question,} we analysed experimental data acquired from cells originating from the adult human brain from Allen Brain Atlas \cite{allen}. In \alain{the} right panels of Fig.~2 we report the histogram of the reversal potential measured experimentally from 200 excitatory (green, upper panel) and 50 inhibitory neurons (red, lower panel). One can see a heterogeneous distribution of the rescaled resting potential $e_L=E_L/\bar{E_L}$, with  $\sigma_I\sim 0.070$ and $\sigma_E\sim 0.056$, grey circle in Fig.~2. \alain{Note} that heterogeneity in inhibitory cells is higher with respect to that in excitatory cells.  \alain{Importantly, these experimental} values fall close to the \alnew{predicted} optimal region of network responsiveness.  \alain{Thus, the model predicts that the optimal heterogeneity matches that found in real neural networks, \alnew{suggesting that this a crucial factor} to understand their responsiveness.}


\begin{figure}
\begin{centering}
\includegraphics[width=0.5\textwidth,clip=true]{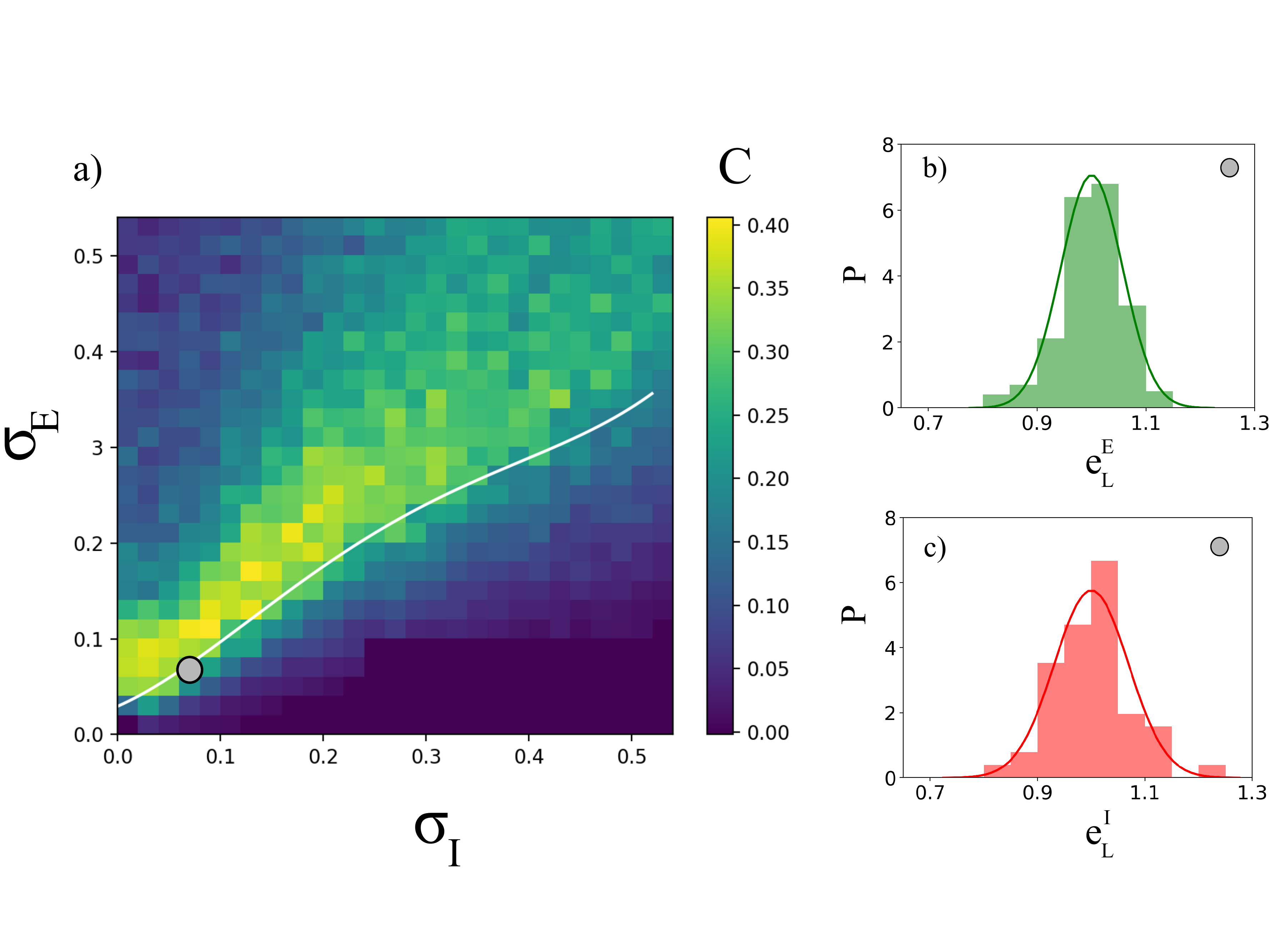}
\end{centering}
\vspace{-20pt}
\caption{\label{fig:2} Panel a) shows the correlation $C$ for different values of $\sigma_E$ and $\sigma_I$. We consider a sinusoidal function for the external rate $\nu_{ext}=\nu_0+A\mathrm{sin}(ft)$, with $A=0.5$Hz, $f=10$Hz and $\nu_0=0.5$Hz. The correlation $C$ is estimated by averaging over 10s. The white line separates the region where the network is characterized by a relatively high  ($r^E>0.1$Hz) spontaneous activity. Panel b) (panel c)) shows the histograms of the resting potential measured over 200 (50) excitatory (inhibitory) neurons in green (red), experimentally measured from cells originating from the adult human brain (data from Allen Brain Atlas \cite{allen} ). The continuous line is a Gaussian distribution with the same standard deviation as measured from data, for which $\sigma_I\sim 0.070$ and $\sigma_E\sim 0.056$.}
\end{figure}


In order to study the mechanism at the origin of the enhanced synchronous population response for heterogeneous inhibitory neurons we developed a mean field approach \alnew{explicitly taking into account diversity}. \alnew{We started from a mean-field model previously introduced for homogeneous neural populations} \cite{el2009master,zerlaut2018modeling,divolo2018biologically}. The corresponding mean field equations describe the dynamics of collective variables, namely excitatory and inhibitory population firing rates, by assuming a Markovian dynamics over a time scale $T\sim \tau_m=15$mS.  We extend this approach to heterogeneous systems by employing a technique, called heterogeneous mean field (HMF), succesfully applied \alnew{previously to model networks with heterogeneous connectivity}
\cite{di2014heterogeneous,burioni2014average}. By performing the thermodynamic limit $N\to\infty$, we consider the network as composed by an infinite amount of classes of neurons, each one characterized by a specific leakage current $E_L^I$. By defining  the transfer function $F_{E^I_L}(r^E,r^I)$ for the class of neurons with leakage reversal $E^I_L$  as neurons' firing activity in function of excitatory (inhibitory) input rate $r^E(r^I)$, the heterogeneous mean field (HMF) equations read:
\begin{align}
&\tau_m\dot r^I = \int dxP(x)F^I_{x}(r^E+\nu_{ext},r^I)-r^I \\
&\tau_m\dot r^E = F^E(r^E+\nu_{ext},r^I,W)-r^E\\
&\tau_w\dot W=-W+br^E+a(V(r^E+\nu_{ext},r^I,W)-E^E_L),
\label{nmf}
\end{align}
where $x=E^I_L$ and $P(x)$ is the distribution of the parameter across the network (a Gaussian distribution with variance $\sigma_I$ in our case) and $W$ the adaptation variable averaged across all neurons.  The estimation of the transfer function $F^E$($F^I$) of excitatory and inhibitory neurons, together with that of the voltage $V(r^E,r^I,W)$, is described in Supplementary Materials. The HMF can be employed following the same steps also for heterogeneity in excitatory neurons. In this case more equations are involved, accounting for the heretogeneous dynamics of adaptation variables $w$ (see Supplementary Materials). By comparing the prediction of the HMF on the response of the network to the eternal stimuli $\nu_{ext}$ we observe a very good agreement, that can be appreciated from the prediction of the response $R$ in function of $\sigma_I$($\sigma_E$), see continuous line in Fig.~1a. 

\begin{figure}[h!]
\begin{centering}
\includegraphics[width=0.55\textwidth,clip=true]{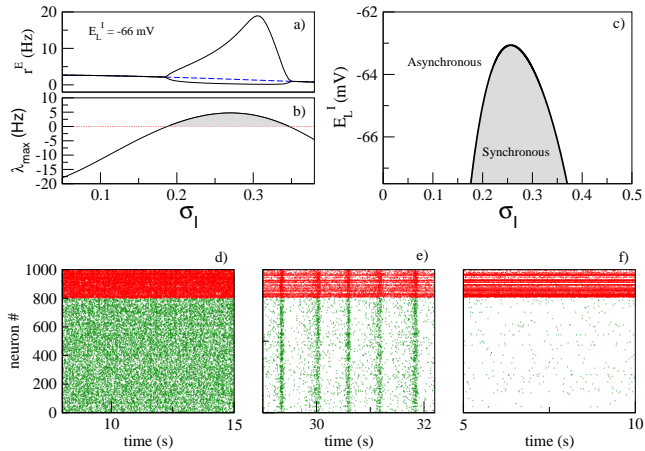}
\end{centering}
\vspace{-20pt}
\caption{\label{fig:4} Upper panels show the analysis of stability of the fixed point from the HMF model. In panel a) we report the stable (unstable) fixed point in continuous (dashed) black (blue) line. Black continuous line shows the maximum and the minimum of oscillations or the firing rate $r^E$. In the lower panel we report, for this fixed point, the maximum of eigenvalues' real parts, $\lambda_{max}$. The grey area shows the region where $\lambda_{max}>0$ and collective oscillations appear. In the right panel c) we report in grey the region where $\lambda_{max}>0$ in function of $\sigma_I$ and the average $E^I_L$. Lower panels show the raster plot from a simulation of $N=10000$ neurons for d) $\sigma_I=0$ , e) $\sigma_I=0.18$ and f) $\sigma_I=0.4$. The constant drive has a frequency $\nu_0=2.5$Hz , $E^I_L=-66$mV and $a=4$nS. }
\end{figure}

The HMF correctly predicts \alnew{the numerical observations of} an optimal amount of heterogeneity in inhibitory neurons for population response, \alnew{but we can use this model to further predict new phenomenon.}  By computing the fixed point ($r^{E^*},r^{I^*},W^*$) we estimate the relative eigenvalues $\{\lambda_i\}$.  Even if the fixed point is always stable, i.e. the  maximum (real part) eigenvalue  $\lambda_{max}<0$, 
we observe that the two first eigenvalues move close to zero in correspondence of the optimal heterogeneity (see Supplementary Materials).
This indicates that the heterogeneity enhances the synchronous population response to external stimuli by modifying the stability proprieties of the asynchronous state. In this perspective, we considered a network setup characterized by a higher recurrent connectivity between excitatory neurons (the probability of connection between two excitatory neuron is $p^{EE}=0.0525$ instead of $p=0.05$). In this case the HMF model predicts that the fixed point is still stable for the homogeneous system ($\lambda_{max}<0$). Nevertheless, the fixed point loses stability by increasing $\sigma_I$ via a Hopf bifurcation (see Fig.~3a) and a sable limit cycle appears. Eventually for even higher values of $\sigma_I$ the fixed point is stable again. This can be observed in  Fig.~3b where we report the maximum eigenvalues $\lambda_{max}$ corresponding to the fixed point ($r^{E^*},r^{I^*},W^*$). We observe a region, marked in grey, with  $\lambda_{max}>0$  where a limit cycle appears. In Fig.~3c we report the critical values for which $\lambda_{max}>0$ in function of $\sigma_I$ and the average $E^I_L$. It is important to notice that the limit cycle is observed only for an heterogeneous system and is not observed in an homogeneous case ($\sigma_I=0$), whatever the modification of the the average excitability of inhibitory neurons. In order to verify these predictions we performed numerical simulations in the spiking network and we observed that, as predicted by the HMF, an  heterogeneous system shows synchronous collective oscillations that are 
not observed in the homogeneous (or too heterogeneous) case, see raster plots in Fig.~3. We have performed a size analyses in order to verify that collective oscillations observed in the network do not disappear in the thermodynamic limit (see Supplementary materials). This result shows that an optimal amount of heterogeneity in inhibitory neurons increases population coherence by synchronizing  the whole network. \alnew{Note that here,} \matteo{the firing activity of single neurons remains irregular even during collective oscillations, as it happens for  sparsely synchronous dynamics in balanced networks  \cite{di2018transition,brunel1999fast}.}

 
In conclusion, we report here three findings.  First, we have found that the heterogeneity of inhibitory neurons, which has been well documented experimentally \cite{Gupta2000,Defelipe2013,Jiang2015,Mihaljevic2019}, optimises the responsiveness of spontaneously active networks to external stimuli.  There appears a resonance peak as a function of the level of heterogeneity. A similar effect of diversity-induced resonance was previously observed in excitable or bistable systems \cite{tessone2006diversity,assisi2005synchrony,lafuerza2010nonuniversal,mejias2012optimal,mejias2014differential}, where heterogeneity in excitatory elements creates active clusters  which were absent in the quiescent  homogeneous system. We showed here that an optimal population response is obtained for heterogeneous excitabilities in both excitatory and inhibitory neurons, via a different mechanism taking place in spontaneously active networks with irregular activity in each neuron.  \alain{Importantly, we found that} the level of heterogeneity measured experimentally corresponds to the resonance peak, which suggests that cortical networks may have naturally evolved towards optimal responsiveness by adjusting their heterogeneity. Moreover, while several studies reported that heterogeneity can enhance coding in uncoupled networks \cite{tripathy2013intermediate,beiran2018coding} and decrease neuronal correlations \cite{padmanabhan2010intrinsic,shamir2006implications,pfeil2016effect}, we report here that a higher input-output population response  is linked to an increased synchronisation observed for an optimal heterogeneity of inhibitory neurons. \alnewnew{The coding capabilities of neural networks
will therefore be largely affected by neuronal heterogeneity, which opens 
interesting perspectives for future studies.}

Second, we designed a mean-field model that explicitly includes heterogeneity, and which can capture this diversity-induced resonance.  This new mean-field formulation keeps track of microscopic complexity, compared to traditional mean-field approaches which implicitly assume homogeneous systems \alnew{and would not predict the correct responsiveness}.

Third, we have shown that neuronal heterogeneity is not only important for responsiveness, but \alnewnew{also} a transition to \alnew{a} sparsely synchronous collective oscillation \alnew{regime} emerges \alnew{by} increasing heterogeneity.  This type of diversity-induced oscillations reminds some aspects found in noise-induced transitions in dynamical systems~\cite{Lefever2006,lindner2004effects}.  Whether the effects of heterogeneity could be considered as analogous to the effect of noise (``quenched noise'') in neural networks is \alnewnew{also} an interesting direction for future studies.
\\
\begin{acknowledgments}

Research funded by the CNRS, the European Community (H2020-785907), \alnew{the ANR (PARADOX) and the ICODE excellence network.} MdV has received financial support by  the ANR Project ERMUNDY (Grant No. 18-CE37-0014-03).

\end{acknowledgments}

  



\begin{thebibliography}{49}%
\makeatletter
\providecommand \@ifxundefined [1]{%
 \@ifx{#1\undefined}
}%
\providecommand \@ifnum [1]{%
 \ifnum #1\expandafter \@firstoftwo
 \else \expandafter \@secondoftwo
 \fi
}%
\providecommand \@ifx [1]{%
 \ifx #1\expandafter \@firstoftwo
 \else \expandafter \@secondoftwo
 \fi
}%
\providecommand \natexlab [1]{#1}%
\providecommand \enquote  [1]{``#1''}%
\providecommand \bibnamefont  [1]{#1}%
\providecommand \bibfnamefont [1]{#1}%
\providecommand \citenamefont [1]{#1}%
\providecommand \href@noop [0]{\@secondoftwo}%
\providecommand \href [0]{\begingroup \@sanitize@url \@href}%
\providecommand \@href[1]{\@@startlink{#1}\@@href}%
\providecommand \@@href[1]{\endgroup#1\@@endlink}%
\providecommand \@sanitize@url [0]{\catcode `\\12\catcode `\$12\catcode
  `\&12\catcode `\#12\catcode `\^12\catcode `\_12\catcode `\%12\relax}%
\providecommand \@@startlink[1]{}%
\providecommand \@@endlink[0]{}%
\providecommand \url  [0]{\begingroup\@sanitize@url \@url }%
\providecommand \@url [1]{\endgroup\@href {#1}{\urlprefix }}%
\providecommand \urlprefix  [0]{URL }%
\providecommand \Eprint [0]{\href }%
\providecommand \doibase [0]{http://dx.doi.org/}%
\providecommand \selectlanguage [0]{\@gobble}%
\providecommand \bibinfo  [0]{\@secondoftwo}%
\providecommand \bibfield  [0]{\@secondoftwo}%
\providecommand \translation [1]{[#1]}%
\providecommand \BibitemOpen [0]{}%
\providecommand \bibitemStop [0]{}%
\providecommand \bibitemNoStop [0]{.\EOS\space}%
\providecommand \EOS [0]{\spacefactor3000\relax}%
\providecommand \BibitemShut  [1]{\csname bibitem#1\endcsname}%
\let\auto@bib@innerbib\@empty
\bibitem [{\citenamefont {Tsodyks}\ \emph {et~al.}(1993)\citenamefont
  {Tsodyks}, \citenamefont {Mitkov},\ and\ \citenamefont
  {Sompolinsky}}]{tsodyks1993pattern}%
  \BibitemOpen
  \bibfield  {author} {\bibinfo {author} {\bibfnamefont {M.}~\bibnamefont
  {Tsodyks}}, \bibinfo {author} {\bibfnamefont {I.}~\bibnamefont {Mitkov}}, \
  and\ \bibinfo {author} {\bibfnamefont {H.}~\bibnamefont {Sompolinsky}},\
  }\href@noop {} {\bibfield  {journal} {\bibinfo  {journal} {Physical review
  letters}\ }\textbf {\bibinfo {volume} {71}},\ \bibinfo {pages} {1280}
  (\bibinfo {year} {1993})}\BibitemShut {NoStop}%
\bibitem [{\citenamefont {Roxin}\ \emph {et~al.}(2011)\citenamefont {Roxin},
  \citenamefont {Brunel}, \citenamefont {Hansel}, \citenamefont {Mongillo},\
  and\ \citenamefont {van Vreeswijk}}]{roxin2011distribution}%
  \BibitemOpen
  \bibfield  {author} {\bibinfo {author} {\bibfnamefont {A.}~\bibnamefont
  {Roxin}}, \bibinfo {author} {\bibfnamefont {N.}~\bibnamefont {Brunel}},
  \bibinfo {author} {\bibfnamefont {D.}~\bibnamefont {Hansel}}, \bibinfo
  {author} {\bibfnamefont {G.}~\bibnamefont {Mongillo}}, \ and\ \bibinfo
  {author} {\bibfnamefont {C.}~\bibnamefont {van Vreeswijk}},\ }\href@noop {}
  {\bibfield  {journal} {\bibinfo  {journal} {Journal of Neuroscience}\
  }\textbf {\bibinfo {volume} {31}},\ \bibinfo {pages} {16217} (\bibinfo {year}
  {2011})}\BibitemShut {NoStop}%
\bibitem [{\citenamefont {Roxin}(2011)}]{roxin2011role}%
  \BibitemOpen
  \bibfield  {author} {\bibinfo {author} {\bibfnamefont {A.}~\bibnamefont
  {Roxin}},\ }\href@noop {} {\bibfield  {journal} {\bibinfo  {journal}
  {Frontiers in computational neuroscience}\ }\textbf {\bibinfo {volume} {5}},\
  \bibinfo {pages} {8} (\bibinfo {year} {2011})}\BibitemShut {NoStop}%
\bibitem [{\citenamefont {Denker}\ \emph {et~al.}(2004)\citenamefont {Denker},
  \citenamefont {Timme}, \citenamefont {Diesmann}, \citenamefont {Wolf},\ and\
  \citenamefont {Geisel}}]{denker2004breaking}%
  \BibitemOpen
  \bibfield  {author} {\bibinfo {author} {\bibfnamefont {M.}~\bibnamefont
  {Denker}}, \bibinfo {author} {\bibfnamefont {M.}~\bibnamefont {Timme}},
  \bibinfo {author} {\bibfnamefont {M.}~\bibnamefont {Diesmann}}, \bibinfo
  {author} {\bibfnamefont {F.}~\bibnamefont {Wolf}}, \ and\ \bibinfo {author}
  {\bibfnamefont {T.}~\bibnamefont {Geisel}},\ }\href@noop {} {\bibfield
  {journal} {\bibinfo  {journal} {Physical review letters}\ }\textbf {\bibinfo
  {volume} {92}},\ \bibinfo {pages} {074103} (\bibinfo {year}
  {2004})}\BibitemShut {NoStop}%
\bibitem [{\citenamefont {Neltner}\ \emph {et~al.}(2000)\citenamefont
  {Neltner}, \citenamefont {Hansel}, \citenamefont {Mato},\ and\ \citenamefont
  {Meunier}}]{neltner2000synchrony}%
  \BibitemOpen
  \bibfield  {author} {\bibinfo {author} {\bibfnamefont {L.}~\bibnamefont
  {Neltner}}, \bibinfo {author} {\bibfnamefont {D.}~\bibnamefont {Hansel}},
  \bibinfo {author} {\bibfnamefont {G.}~\bibnamefont {Mato}}, \ and\ \bibinfo
  {author} {\bibfnamefont {C.}~\bibnamefont {Meunier}},\ }\href@noop {}
  {\bibfield  {journal} {\bibinfo  {journal} {Neural computation}\ }\textbf
  {\bibinfo {volume} {12}},\ \bibinfo {pages} {1607} (\bibinfo {year}
  {2000})}\BibitemShut {NoStop}%
\bibitem [{\citenamefont {Golomb}\ and\ \citenamefont
  {Rinzel}(1993)}]{golomb1993dynamics}%
  \BibitemOpen
  \bibfield  {author} {\bibinfo {author} {\bibfnamefont {D.}~\bibnamefont
  {Golomb}}\ and\ \bibinfo {author} {\bibfnamefont {J.}~\bibnamefont
  {Rinzel}},\ }\href@noop {} {\bibfield  {journal} {\bibinfo  {journal}
  {Physical review E}\ }\textbf {\bibinfo {volume} {48}},\ \bibinfo {pages}
  {4810} (\bibinfo {year} {1993})}\BibitemShut {NoStop}%
\bibitem [{\citenamefont {Zerlaut}\ \emph {et~al.}(2016)\citenamefont
  {Zerlaut}, \citenamefont {Tele{\'n}czuk}, \citenamefont {Deleuze},
  \citenamefont {Bal}, \citenamefont {Ouanounou},\ and\ \citenamefont
  {Destexhe}}]{zerlaut2016heterogeneous}%
  \BibitemOpen
  \bibfield  {author} {\bibinfo {author} {\bibfnamefont {Y.}~\bibnamefont
  {Zerlaut}}, \bibinfo {author} {\bibfnamefont {B.}~\bibnamefont
  {Tele{\'n}czuk}}, \bibinfo {author} {\bibfnamefont {C.}~\bibnamefont
  {Deleuze}}, \bibinfo {author} {\bibfnamefont {T.}~\bibnamefont {Bal}},
  \bibinfo {author} {\bibfnamefont {G.}~\bibnamefont {Ouanounou}}, \ and\
  \bibinfo {author} {\bibfnamefont {A.}~\bibnamefont {Destexhe}},\ }\href@noop
  {} {\bibfield  {journal} {\bibinfo  {journal} {The Journal of physiology}\
  }\textbf {\bibinfo {volume} {594}},\ \bibinfo {pages} {3791} (\bibinfo {year}
  {2016})}\BibitemShut {NoStop}%
\bibitem [{\citenamefont {Tseng}\ and\ \citenamefont
  {Prince}(1993)}]{tseng1993heterogeneity}%
  \BibitemOpen
  \bibfield  {author} {\bibinfo {author} {\bibfnamefont {G.-F.}\ \bibnamefont
  {Tseng}}\ and\ \bibinfo {author} {\bibfnamefont {D.~A.}\ \bibnamefont
  {Prince}},\ }\href@noop {} {\bibfield  {journal} {\bibinfo  {journal}
  {Journal of Comparative Neurology}\ }\textbf {\bibinfo {volume} {335}},\
  \bibinfo {pages} {92} (\bibinfo {year} {1993})}\BibitemShut {NoStop}%
\bibitem [{\citenamefont {Pospischil}\ \emph {et~al.}(2008)\citenamefont
  {Pospischil}, \citenamefont {Toledo-Rodriguez}, \citenamefont {Monier},
  \citenamefont {Piwkowska}, \citenamefont {Bal}, \citenamefont {Fr{\'e}gnac},
  \citenamefont {Markram},\ and\ \citenamefont
  {Destexhe}}]{pospischil2008minimal}%
  \BibitemOpen
  \bibfield  {author} {\bibinfo {author} {\bibfnamefont {M.}~\bibnamefont
  {Pospischil}}, \bibinfo {author} {\bibfnamefont {M.}~\bibnamefont
  {Toledo-Rodriguez}}, \bibinfo {author} {\bibfnamefont {C.}~\bibnamefont
  {Monier}}, \bibinfo {author} {\bibfnamefont {Z.}~\bibnamefont {Piwkowska}},
  \bibinfo {author} {\bibfnamefont {T.}~\bibnamefont {Bal}}, \bibinfo {author}
  {\bibfnamefont {Y.}~\bibnamefont {Fr{\'e}gnac}}, \bibinfo {author}
  {\bibfnamefont {H.}~\bibnamefont {Markram}}, \ and\ \bibinfo {author}
  {\bibfnamefont {A.}~\bibnamefont {Destexhe}},\ }\href@noop {} {\bibfield
  {journal} {\bibinfo  {journal} {Biological cybernetics}\ }\textbf {\bibinfo
  {volume} {99}},\ \bibinfo {pages} {427} (\bibinfo {year} {2008})}\BibitemShut
  {NoStop}%
\bibitem [{\citenamefont {Sharpee}(2014)}]{sharpee2014toward}%
  \BibitemOpen
  \bibfield  {author} {\bibinfo {author} {\bibfnamefont {T.~O.}\ \bibnamefont
  {Sharpee}},\ }\href@noop {} {\bibfield  {journal} {\bibinfo  {journal}
  {Neuron}\ }\textbf {\bibinfo {volume} {83}},\ \bibinfo {pages} {1329}
  (\bibinfo {year} {2014})}\BibitemShut {NoStop}%
\bibitem [{\citenamefont {Landau}\ \emph {et~al.}(2016)\citenamefont {Landau},
  \citenamefont {Egger}, \citenamefont {Dercksen}, \citenamefont
  {Oberlaender},\ and\ \citenamefont {Sompolinsky}}]{landau2016impact}%
  \BibitemOpen
  \bibfield  {author} {\bibinfo {author} {\bibfnamefont {I.~D.}\ \bibnamefont
  {Landau}}, \bibinfo {author} {\bibfnamefont {R.}~\bibnamefont {Egger}},
  \bibinfo {author} {\bibfnamefont {V.~J.}\ \bibnamefont {Dercksen}}, \bibinfo
  {author} {\bibfnamefont {M.}~\bibnamefont {Oberlaender}}, \ and\ \bibinfo
  {author} {\bibfnamefont {H.}~\bibnamefont {Sompolinsky}},\ }\href@noop {}
  {\bibfield  {journal} {\bibinfo  {journal} {Neuron}\ }\textbf {\bibinfo
  {volume} {92}},\ \bibinfo {pages} {1106} (\bibinfo {year}
  {2016})}\BibitemShut {NoStop}%
\bibitem [{\citenamefont {Gupta}\ \emph {et~al.}(2000)\citenamefont {Gupta},
  \citenamefont {Wang},\ and\ \citenamefont {Markram}}]{Gupta2000}%
  \BibitemOpen
  \bibfield  {author} {\bibinfo {author} {\bibfnamefont {A.}~\bibnamefont
  {Gupta}}, \bibinfo {author} {\bibfnamefont {Y.}~\bibnamefont {Wang}}, \ and\
  \bibinfo {author} {\bibfnamefont {H.}~\bibnamefont {Markram}},\ }\href@noop
  {} {\bibfield  {journal} {\bibinfo  {journal} {Science}\ }\textbf {\bibinfo
  {volume} {287}},\ \bibinfo {pages} {273} (\bibinfo {year}
  {2000})}\BibitemShut {NoStop}%
\bibitem [{\citenamefont {DeFelipe}\ \emph {et~al.}(2013)\citenamefont
  {DeFelipe}, \citenamefont {L?pez-Cruz}, \citenamefont {Benavides-Piccione},
  \citenamefont {Bielza}, \citenamefont {Larra?aga}, \citenamefont {Anderson},
  \citenamefont {Burkhalter}, \citenamefont {Cauli}, \citenamefont {Fair?n},
  \citenamefont {Feldmeyer}, \citenamefont {Fishell}, \citenamefont
  {Fitzpatrick}, \citenamefont {Freund}, \citenamefont {Gonz?lez-Burgos},
  \citenamefont {Hestrin}, \citenamefont {Hill}, \citenamefont {Hof},
  \citenamefont {Huang}, \citenamefont {Jones}, \citenamefont {Kawaguchi},
  \citenamefont {Kisv?rday}, \citenamefont {Kubota}, \citenamefont {Lewis},
  \citenamefont {Mar?n}, \citenamefont {Markram}, \citenamefont {McBain},
  \citenamefont {Meyer}, \citenamefont {Monyer}, \citenamefont {Nelson},
  \citenamefont {Rockland}, \citenamefont {Rossier}, \citenamefont
  {Rubenstein}, \citenamefont {Rudy}, \citenamefont {Scanziani}, \citenamefont
  {Shepherd}, \citenamefont {Sherwood}, \citenamefont {Staiger}, \citenamefont
  {Tam?s}, \citenamefont {Thomson}, \citenamefont {Wang}, \citenamefont
  {Yuste},\ and\ \citenamefont {Ascoli}}]{Defelipe2013}%
  \BibitemOpen
  \bibfield  {author} {\bibinfo {author} {\bibfnamefont {J.}~\bibnamefont
  {DeFelipe}}, \bibinfo {author} {\bibfnamefont {P.~L.}\ \bibnamefont
  {L?pez-Cruz}}, \bibinfo {author} {\bibfnamefont {R.}~\bibnamefont
  {Benavides-Piccione}}, \bibinfo {author} {\bibfnamefont {C.}~\bibnamefont
  {Bielza}}, \bibinfo {author} {\bibfnamefont {P.}~\bibnamefont {Larra?aga}},
  \bibinfo {author} {\bibfnamefont {S.}~\bibnamefont {Anderson}}, \bibinfo
  {author} {\bibfnamefont {A.}~\bibnamefont {Burkhalter}}, \bibinfo {author}
  {\bibfnamefont {B.}~\bibnamefont {Cauli}}, \bibinfo {author} {\bibfnamefont
  {A.}~\bibnamefont {Fair?n}}, \bibinfo {author} {\bibfnamefont
  {D.}~\bibnamefont {Feldmeyer}}, \bibinfo {author} {\bibfnamefont
  {G.}~\bibnamefont {Fishell}}, \bibinfo {author} {\bibfnamefont
  {D.}~\bibnamefont {Fitzpatrick}}, \bibinfo {author} {\bibfnamefont {T.~F.}\
  \bibnamefont {Freund}}, \bibinfo {author} {\bibfnamefont {G.}~\bibnamefont
  {Gonz?lez-Burgos}}, \bibinfo {author} {\bibfnamefont {S.}~\bibnamefont
  {Hestrin}}, \bibinfo {author} {\bibfnamefont {S.}~\bibnamefont {Hill}},
  \bibinfo {author} {\bibfnamefont {P.~R.}\ \bibnamefont {Hof}}, \bibinfo
  {author} {\bibfnamefont {J.}~\bibnamefont {Huang}}, \bibinfo {author}
  {\bibfnamefont {E.~G.}\ \bibnamefont {Jones}}, \bibinfo {author}
  {\bibfnamefont {Y.}~\bibnamefont {Kawaguchi}}, \bibinfo {author}
  {\bibfnamefont {Z.}~\bibnamefont {Kisv?rday}}, \bibinfo {author}
  {\bibfnamefont {Y.}~\bibnamefont {Kubota}}, \bibinfo {author} {\bibfnamefont
  {D.~A.}\ \bibnamefont {Lewis}}, \bibinfo {author} {\bibfnamefont
  {O.}~\bibnamefont {Mar?n}}, \bibinfo {author} {\bibfnamefont
  {H.}~\bibnamefont {Markram}}, \bibinfo {author} {\bibfnamefont {C.~J.}\
  \bibnamefont {McBain}}, \bibinfo {author} {\bibfnamefont {H.~S.}\
  \bibnamefont {Meyer}}, \bibinfo {author} {\bibfnamefont {H.}~\bibnamefont
  {Monyer}}, \bibinfo {author} {\bibfnamefont {S.~B.}\ \bibnamefont {Nelson}},
  \bibinfo {author} {\bibfnamefont {K.}~\bibnamefont {Rockland}}, \bibinfo
  {author} {\bibfnamefont {J.}~\bibnamefont {Rossier}}, \bibinfo {author}
  {\bibfnamefont {J.~L.}\ \bibnamefont {Rubenstein}}, \bibinfo {author}
  {\bibfnamefont {B.}~\bibnamefont {Rudy}}, \bibinfo {author} {\bibfnamefont
  {M.}~\bibnamefont {Scanziani}}, \bibinfo {author} {\bibfnamefont {G.~M.}\
  \bibnamefont {Shepherd}}, \bibinfo {author} {\bibfnamefont {C.~C.}\
  \bibnamefont {Sherwood}}, \bibinfo {author} {\bibfnamefont {J.~F.}\
  \bibnamefont {Staiger}}, \bibinfo {author} {\bibfnamefont {G.}~\bibnamefont
  {Tam?s}}, \bibinfo {author} {\bibfnamefont {A.}~\bibnamefont {Thomson}},
  \bibinfo {author} {\bibfnamefont {Y.}~\bibnamefont {Wang}}, \bibinfo {author}
  {\bibfnamefont {R.}~\bibnamefont {Yuste}}, \ and\ \bibinfo {author}
  {\bibfnamefont {G.~A.}\ \bibnamefont {Ascoli}},\ }\href@noop {} {\bibfield
  {journal} {\bibinfo  {journal} {Nat. Rev. Neurosci.}\ }\textbf {\bibinfo
  {volume} {14}},\ \bibinfo {pages} {202} (\bibinfo {year} {2013})}\BibitemShut
  {NoStop}%
\bibitem [{\citenamefont {Jiang}\ \emph {et~al.}(2015)\citenamefont {Jiang},
  \citenamefont {Shen}, \citenamefont {Cadwell}, \citenamefont {Berens},
  \citenamefont {Sinz}, \citenamefont {Ecker}, \citenamefont {Patel},\ and\
  \citenamefont {Tolias}}]{Jiang2015}%
  \BibitemOpen
  \bibfield  {author} {\bibinfo {author} {\bibfnamefont {X.}~\bibnamefont
  {Jiang}}, \bibinfo {author} {\bibfnamefont {S.}~\bibnamefont {Shen}},
  \bibinfo {author} {\bibfnamefont {C.~R.}\ \bibnamefont {Cadwell}}, \bibinfo
  {author} {\bibfnamefont {P.}~\bibnamefont {Berens}}, \bibinfo {author}
  {\bibfnamefont {F.}~\bibnamefont {Sinz}}, \bibinfo {author} {\bibfnamefont
  {A.~S.}\ \bibnamefont {Ecker}}, \bibinfo {author} {\bibfnamefont
  {S.}~\bibnamefont {Patel}}, \ and\ \bibinfo {author} {\bibfnamefont {A.~S.}\
  \bibnamefont {Tolias}},\ }\href@noop {} {\bibfield  {journal} {\bibinfo
  {journal} {Science}\ }\textbf {\bibinfo {volume} {350}},\ \bibinfo {pages}
  {aac9462} (\bibinfo {year} {2015})}\BibitemShut {NoStop}%
\bibitem [{\citenamefont {Mihaljevic}\ \emph {et~al.}(2019)\citenamefont
  {Mihaljevic}, \citenamefont {Benavides-Piccione}, \citenamefont {Bielza},
  \citenamefont {Larra?aga},\ and\ \citenamefont {DeFelipe}}]{Mihaljevic2019}%
  \BibitemOpen
  \bibfield  {author} {\bibinfo {author} {\bibfnamefont {B.}~\bibnamefont
  {Mihaljevic}}, \bibinfo {author} {\bibfnamefont {R.}~\bibnamefont
  {Benavides-Piccione}}, \bibinfo {author} {\bibfnamefont {C.}~\bibnamefont
  {Bielza}}, \bibinfo {author} {\bibfnamefont {P.}~\bibnamefont {Larra?aga}}, \
  and\ \bibinfo {author} {\bibfnamefont {J.}~\bibnamefont {DeFelipe}},\
  }\href@noop {} {\bibfield  {journal} {\bibinfo  {journal} {Sci Data}\
  }\textbf {\bibinfo {volume} {6}},\ \bibinfo {pages} {221} (\bibinfo {year}
  {2019})}\BibitemShut {NoStop}%
\bibitem [{\citenamefont {Chow}(1998)}]{chow1998phase}%
  \BibitemOpen
  \bibfield  {author} {\bibinfo {author} {\bibfnamefont {C.~C.}\ \bibnamefont
  {Chow}},\ }\href@noop {} {\bibfield  {journal} {\bibinfo  {journal} {Physica
  D: Nonlinear Phenomena}\ }\textbf {\bibinfo {volume} {118}},\ \bibinfo
  {pages} {343} (\bibinfo {year} {1998})}\BibitemShut {NoStop}%
\bibitem [{\citenamefont {White}\ \emph {et~al.}(1998)\citenamefont {White},
  \citenamefont {Chow}, \citenamefont {Rit}, \citenamefont {Soto-Trevi{\~n}o},\
  and\ \citenamefont {Kopell}}]{white1998synchronization}%
  \BibitemOpen
  \bibfield  {author} {\bibinfo {author} {\bibfnamefont {J.~A.}\ \bibnamefont
  {White}}, \bibinfo {author} {\bibfnamefont {C.~C.}\ \bibnamefont {Chow}},
  \bibinfo {author} {\bibfnamefont {J.}~\bibnamefont {Rit}}, \bibinfo {author}
  {\bibfnamefont {C.}~\bibnamefont {Soto-Trevi{\~n}o}}, \ and\ \bibinfo
  {author} {\bibfnamefont {N.}~\bibnamefont {Kopell}},\ }\href@noop {}
  {\bibfield  {journal} {\bibinfo  {journal} {Journal of computational
  neuroscience}\ }\textbf {\bibinfo {volume} {5}},\ \bibinfo {pages} {5}
  (\bibinfo {year} {1998})}\BibitemShut {NoStop}%
\bibitem [{\citenamefont {Traub}\ \emph {et~al.}(2005)\citenamefont {Traub},
  \citenamefont {Contreras}, \citenamefont {Cunningham}, \citenamefont
  {Murray}, \citenamefont {LeBeau}, \citenamefont {Roopun}, \citenamefont
  {Bibbig}, \citenamefont {Wilent}, \citenamefont {Higley},\ and\ \citenamefont
  {Whittington}}]{traub2005single}%
  \BibitemOpen
  \bibfield  {author} {\bibinfo {author} {\bibfnamefont {R.~D.}\ \bibnamefont
  {Traub}}, \bibinfo {author} {\bibfnamefont {D.}~\bibnamefont {Contreras}},
  \bibinfo {author} {\bibfnamefont {M.~O.}\ \bibnamefont {Cunningham}},
  \bibinfo {author} {\bibfnamefont {H.}~\bibnamefont {Murray}}, \bibinfo
  {author} {\bibfnamefont {F.~E.}\ \bibnamefont {LeBeau}}, \bibinfo {author}
  {\bibfnamefont {A.}~\bibnamefont {Roopun}}, \bibinfo {author} {\bibfnamefont
  {A.}~\bibnamefont {Bibbig}}, \bibinfo {author} {\bibfnamefont {W.~B.}\
  \bibnamefont {Wilent}}, \bibinfo {author} {\bibfnamefont {M.~J.}\
  \bibnamefont {Higley}}, \ and\ \bibinfo {author} {\bibfnamefont {M.~A.}\
  \bibnamefont {Whittington}},\ }\href@noop {} {\bibfield  {journal} {\bibinfo
  {journal} {Journal of neurophysiology}\ }\textbf {\bibinfo {volume} {93}},\
  \bibinfo {pages} {2194} (\bibinfo {year} {2005})}\BibitemShut {NoStop}%
\bibitem [{\citenamefont {Wang}\ and\ \citenamefont
  {Buzs{\'a}ki}(1996)}]{wang1996gamma}%
  \BibitemOpen
  \bibfield  {author} {\bibinfo {author} {\bibfnamefont {X.-J.}\ \bibnamefont
  {Wang}}\ and\ \bibinfo {author} {\bibfnamefont {G.}~\bibnamefont
  {Buzs{\'a}ki}},\ }\href@noop {} {\bibfield  {journal} {\bibinfo  {journal}
  {Journal of neuroscience}\ }\textbf {\bibinfo {volume} {16}},\ \bibinfo
  {pages} {6402} (\bibinfo {year} {1996})}\BibitemShut {NoStop}%
\bibitem [{\citenamefont {Tiesinga}\ and\ \citenamefont
  {Jos{\'e}}(2000)}]{tiesinga2000robust}%
  \BibitemOpen
  \bibfield  {author} {\bibinfo {author} {\bibfnamefont {P.}~\bibnamefont
  {Tiesinga}}\ and\ \bibinfo {author} {\bibfnamefont {J.~V.}\ \bibnamefont
  {Jos{\'e}}},\ }\href@noop {} {\bibfield  {journal} {\bibinfo  {journal}
  {Network: Computation in Neural Systems}\ }\textbf {\bibinfo {volume} {11}},\
  \bibinfo {pages} {1} (\bibinfo {year} {2000})}\BibitemShut {NoStop}%
\bibitem [{\citenamefont {Devalle}\ \emph {et~al.}(2018)\citenamefont
  {Devalle}, \citenamefont {Montbri{\'o}},\ and\ \citenamefont
  {Paz{\'o}}}]{devalle2018dynamics}%
  \BibitemOpen
  \bibfield  {author} {\bibinfo {author} {\bibfnamefont {F.}~\bibnamefont
  {Devalle}}, \bibinfo {author} {\bibfnamefont {E.}~\bibnamefont
  {Montbri{\'o}}}, \ and\ \bibinfo {author} {\bibfnamefont {D.}~\bibnamefont
  {Paz{\'o}}},\ }\href@noop {} {\bibfield  {journal} {\bibinfo  {journal}
  {Physical Review E}\ }\textbf {\bibinfo {volume} {98}},\ \bibinfo {pages}
  {042214} (\bibinfo {year} {2018})}\BibitemShut {NoStop}%
\bibitem [{\citenamefont {Luccioli}\ \emph {et~al.}(2019)\citenamefont
  {Luccioli}, \citenamefont {Angulo-Garcia},\ and\ \citenamefont
  {Torcini}}]{luccioli2019neural}%
  \BibitemOpen
  \bibfield  {author} {\bibinfo {author} {\bibfnamefont {S.}~\bibnamefont
  {Luccioli}}, \bibinfo {author} {\bibfnamefont {D.}~\bibnamefont
  {Angulo-Garcia}}, \ and\ \bibinfo {author} {\bibfnamefont {A.}~\bibnamefont
  {Torcini}},\ }\href@noop {} {\bibfield  {journal} {\bibinfo  {journal}
  {Physical Review E}\ }\textbf {\bibinfo {volume} {99}},\ \bibinfo {pages}
  {052412} (\bibinfo {year} {2019})}\BibitemShut {NoStop}%
\bibitem [{\citenamefont {Gray}(1994)}]{gray1994synchronous}%
  \BibitemOpen
  \bibfield  {author} {\bibinfo {author} {\bibfnamefont {C.~M.}\ \bibnamefont
  {Gray}},\ }\href@noop {} {\bibfield  {journal} {\bibinfo  {journal} {Journal
  of computational neuroscience}\ }\textbf {\bibinfo {volume} {1}},\ \bibinfo
  {pages} {11} (\bibinfo {year} {1994})}\BibitemShut {NoStop}%
\bibitem [{\citenamefont {Llinas}\ and\ \citenamefont
  {Ribary}(1993)}]{llinas1993coherent}%
  \BibitemOpen
  \bibfield  {author} {\bibinfo {author} {\bibfnamefont {R.}~\bibnamefont
  {Llinas}}\ and\ \bibinfo {author} {\bibfnamefont {U.}~\bibnamefont
  {Ribary}},\ }\href@noop {} {\bibfield  {journal} {\bibinfo  {journal}
  {Proceedings of the National Academy of Sciences}\ }\textbf {\bibinfo
  {volume} {90}},\ \bibinfo {pages} {2078} (\bibinfo {year}
  {1993})}\BibitemShut {NoStop}%
\bibitem [{\citenamefont {Brette}\ and\ \citenamefont
  {Gerstner}(2005)}]{brette2005adaptive}%
  \BibitemOpen
  \bibfield  {author} {\bibinfo {author} {\bibfnamefont {R.}~\bibnamefont
  {Brette}}\ and\ \bibinfo {author} {\bibfnamefont {W.}~\bibnamefont
  {Gerstner}},\ }\href@noop {} {\bibfield  {journal} {\bibinfo  {journal}
  {Journal of Neurophysiology}\ }\textbf {\bibinfo {volume} {94}},\ \bibinfo
  {pages} {3637} (\bibinfo {year} {2005})}\BibitemShut {NoStop}%
\bibitem [{\citenamefont {Monteforte}\ and\ \citenamefont
  {Wolf}(2010)}]{monteforte2010dynamical}%
  \BibitemOpen
  \bibfield  {author} {\bibinfo {author} {\bibfnamefont {M.}~\bibnamefont
  {Monteforte}}\ and\ \bibinfo {author} {\bibfnamefont {F.}~\bibnamefont
  {Wolf}},\ }\href@noop {} {\bibfield  {journal} {\bibinfo  {journal} {Physical
  review letters}\ }\textbf {\bibinfo {volume} {105}},\ \bibinfo {pages}
  {268104} (\bibinfo {year} {2010})}\BibitemShut {NoStop}%
\bibitem [{\citenamefont {Van~Vreeswijk}\ and\ \citenamefont
  {Sompolinsky}(1996)}]{van1996chaos}%
  \BibitemOpen
  \bibfield  {author} {\bibinfo {author} {\bibfnamefont {C.}~\bibnamefont
  {Van~Vreeswijk}}\ and\ \bibinfo {author} {\bibfnamefont {H.}~\bibnamefont
  {Sompolinsky}},\ }\href@noop {} {\bibfield  {journal} {\bibinfo  {journal}
  {Science}\ }\textbf {\bibinfo {volume} {274}},\ \bibinfo {pages} {1724}
  (\bibinfo {year} {1996})}\BibitemShut {NoStop}%
\bibitem [{\citenamefont {Destexhe}\ \emph {et~al.}(2003)\citenamefont
  {Destexhe}, \citenamefont {Rudolph},\ and\ \citenamefont
  {Par{\'e}}}]{destexhe2003high}%
  \BibitemOpen
  \bibfield  {author} {\bibinfo {author} {\bibfnamefont {A.}~\bibnamefont
  {Destexhe}}, \bibinfo {author} {\bibfnamefont {M.}~\bibnamefont {Rudolph}}, \
  and\ \bibinfo {author} {\bibfnamefont {D.}~\bibnamefont {Par{\'e}}},\
  }\href@noop {} {\bibfield  {journal} {\bibinfo  {journal} {Nature reviews
  neuroscience}\ }\textbf {\bibinfo {volume} {4}},\ \bibinfo {pages} {739}
  (\bibinfo {year} {2003})}\BibitemShut {NoStop}%
\bibitem [{\citenamefont {Mejias}\ and\ \citenamefont
  {Longtin}(2012)}]{mejias2012optimal}%
  \BibitemOpen
  \bibfield  {author} {\bibinfo {author} {\bibfnamefont {J.}~\bibnamefont
  {Mejias}}\ and\ \bibinfo {author} {\bibfnamefont {A.}~\bibnamefont
  {Longtin}},\ }\href@noop {} {\bibfield  {journal} {\bibinfo  {journal}
  {Physical Review Letters}\ }\textbf {\bibinfo {volume} {108}},\ \bibinfo
  {pages} {228102} (\bibinfo {year} {2012})}\BibitemShut {NoStop}%
\bibitem [{\citenamefont {Gollo}\ \emph {et~al.}(2016)\citenamefont {Gollo},
  \citenamefont {Copelli},\ and\ \citenamefont {Roberts}}]{gollo2016diversity}%
  \BibitemOpen
  \bibfield  {author} {\bibinfo {author} {\bibfnamefont {L.~L.}\ \bibnamefont
  {Gollo}}, \bibinfo {author} {\bibfnamefont {M.}~\bibnamefont {Copelli}}, \
  and\ \bibinfo {author} {\bibfnamefont {J.~A.}\ \bibnamefont {Roberts}},\
  }\href@noop {} {\bibfield  {journal} {\bibinfo  {journal} {PeerJ}\ }\textbf
  {\bibinfo {volume} {4}},\ \bibinfo {pages} {e1912} (\bibinfo {year}
  {2016})}\BibitemShut {NoStop}%
\bibitem [{all()}]{allen}%
  \BibitemOpen
  \href@noop {} {\bibinfo  {journal} {2015 Allen Cell Types Database, Available
  from: https://celltypes.brain-map.org/overview}\ }\BibitemShut {NoStop}%
\bibitem [{\citenamefont {El~Boustani}\ and\ \citenamefont
  {Destexhe}(2009)}]{el2009master}%
  \BibitemOpen
\bibfield  {journal} {  }\bibfield  {author} {\bibinfo {author} {\bibfnamefont
  {S.}~\bibnamefont {El~Boustani}}\ and\ \bibinfo {author} {\bibfnamefont
  {A.}~\bibnamefont {Destexhe}},\ }\href@noop {} {\bibfield  {journal}
  {\bibinfo  {journal} {Neural computation}\ }\textbf {\bibinfo {volume}
  {21}},\ \bibinfo {pages} {46} (\bibinfo {year} {2009})}\BibitemShut {NoStop}%
\bibitem [{\citenamefont {Zerlaut}\ \emph {et~al.}(2018)\citenamefont
  {Zerlaut}, \citenamefont {Chemla}, \citenamefont {Chavane},\ and\
  \citenamefont {Destexhe}}]{zerlaut2018modeling}%
  \BibitemOpen
  \bibfield  {author} {\bibinfo {author} {\bibfnamefont {Y.}~\bibnamefont
  {Zerlaut}}, \bibinfo {author} {\bibfnamefont {S.}~\bibnamefont {Chemla}},
  \bibinfo {author} {\bibfnamefont {F.}~\bibnamefont {Chavane}}, \ and\
  \bibinfo {author} {\bibfnamefont {A.}~\bibnamefont {Destexhe}},\ }\href@noop
  {} {\bibfield  {journal} {\bibinfo  {journal} {Journal of computational
  neuroscience}\ }\textbf {\bibinfo {volume} {44}},\ \bibinfo {pages} {45}
  (\bibinfo {year} {2018})}\BibitemShut {NoStop}%
\bibitem [{\citenamefont {di~Volo}\ \emph {et~al.}(2019)\citenamefont
  {di~Volo}, \citenamefont {Romagnoni}, \citenamefont {Capone},\ and\
  \citenamefont {Destexhe}}]{divolo2018biologically}%
  \BibitemOpen
  \bibfield  {author} {\bibinfo {author} {\bibfnamefont {M.}~\bibnamefont
  {di~Volo}}, \bibinfo {author} {\bibfnamefont {A.}~\bibnamefont {Romagnoni}},
  \bibinfo {author} {\bibfnamefont {C.}~\bibnamefont {Capone}}, \ and\ \bibinfo
  {author} {\bibfnamefont {A.}~\bibnamefont {Destexhe}},\ }\href@noop {}
  {\bibfield  {journal} {\bibinfo  {journal} {Neural Computation}\ }\textbf
  {\bibinfo {volume} {31}},\ \bibinfo {pages} {653} (\bibinfo {year}
  {2019})}\BibitemShut {NoStop}%
\bibitem [{\citenamefont {di~Volo}\ \emph {et~al.}(2014)\citenamefont
  {di~Volo}, \citenamefont {Burioni}, \citenamefont {Casartelli}, \citenamefont
  {Livi},\ and\ \citenamefont {Vezzani}}]{di2014heterogeneous}%
  \BibitemOpen
  \bibfield  {author} {\bibinfo {author} {\bibfnamefont {M.}~\bibnamefont
  {di~Volo}}, \bibinfo {author} {\bibfnamefont {R.}~\bibnamefont {Burioni}},
  \bibinfo {author} {\bibfnamefont {M.}~\bibnamefont {Casartelli}}, \bibinfo
  {author} {\bibfnamefont {R.}~\bibnamefont {Livi}}, \ and\ \bibinfo {author}
  {\bibfnamefont {A.}~\bibnamefont {Vezzani}},\ }\href@noop {} {\bibfield
  {journal} {\bibinfo  {journal} {Physical Review E}\ }\textbf {\bibinfo
  {volume} {90}},\ \bibinfo {pages} {022811} (\bibinfo {year}
  {2014})}\BibitemShut {NoStop}%
\bibitem [{\citenamefont {Burioni}\ \emph {et~al.}(2014)\citenamefont
  {Burioni}, \citenamefont {Casartelli}, \citenamefont {Di~Volo}, \citenamefont
  {Livi},\ and\ \citenamefont {Vezzani}}]{burioni2014average}%
  \BibitemOpen
  \bibfield  {author} {\bibinfo {author} {\bibfnamefont {R.}~\bibnamefont
  {Burioni}}, \bibinfo {author} {\bibfnamefont {M.}~\bibnamefont {Casartelli}},
  \bibinfo {author} {\bibfnamefont {M.}~\bibnamefont {Di~Volo}}, \bibinfo
  {author} {\bibfnamefont {R.}~\bibnamefont {Livi}}, \ and\ \bibinfo {author}
  {\bibfnamefont {A.}~\bibnamefont {Vezzani}},\ }\href@noop {} {\bibfield
  {journal} {\bibinfo  {journal} {Scientific reports}\ }\textbf {\bibinfo
  {volume} {4}},\ \bibinfo {pages} {1} (\bibinfo {year} {2014})}\BibitemShut
  {NoStop}%
\bibitem [{\citenamefont {di~Volo}\ and\ \citenamefont
  {Torcini}(2018)}]{di2018transition}%
  \BibitemOpen
  \bibfield  {author} {\bibinfo {author} {\bibfnamefont {M.}~\bibnamefont
  {di~Volo}}\ and\ \bibinfo {author} {\bibfnamefont {A.}~\bibnamefont
  {Torcini}},\ }\href@noop {} {\bibfield  {journal} {\bibinfo  {journal}
  {Physical review letters}\ }\textbf {\bibinfo {volume} {121}},\ \bibinfo
  {pages} {128301} (\bibinfo {year} {2018})}\BibitemShut {NoStop}%
\bibitem [{\citenamefont {Brunel}\ and\ \citenamefont
  {Hakim}(1999)}]{brunel1999fast}%
  \BibitemOpen
  \bibfield  {author} {\bibinfo {author} {\bibfnamefont {N.}~\bibnamefont
  {Brunel}}\ and\ \bibinfo {author} {\bibfnamefont {V.}~\bibnamefont {Hakim}},\
  }\href@noop {} {\bibfield  {journal} {\bibinfo  {journal} {Neural
  computation}\ }\textbf {\bibinfo {volume} {11}},\ \bibinfo {pages} {1621}
  (\bibinfo {year} {1999})}\BibitemShut {NoStop}%
\bibitem [{\citenamefont {Tessone}\ \emph {et~al.}(2006)\citenamefont
  {Tessone}, \citenamefont {Mirasso}, \citenamefont {Toral},\ and\
  \citenamefont {Gunton}}]{tessone2006diversity}%
  \BibitemOpen
  \bibfield  {author} {\bibinfo {author} {\bibfnamefont {C.~J.}\ \bibnamefont
  {Tessone}}, \bibinfo {author} {\bibfnamefont {C.~R.}\ \bibnamefont
  {Mirasso}}, \bibinfo {author} {\bibfnamefont {R.}~\bibnamefont {Toral}}, \
  and\ \bibinfo {author} {\bibfnamefont {J.~D.}\ \bibnamefont {Gunton}},\
  }\href@noop {} {\bibfield  {journal} {\bibinfo  {journal} {Physical review
  letters}\ }\textbf {\bibinfo {volume} {97}},\ \bibinfo {pages} {194101}
  (\bibinfo {year} {2006})}\BibitemShut {NoStop}%
\bibitem [{\citenamefont {Assisi}\ \emph {et~al.}(2005)\citenamefont {Assisi},
  \citenamefont {Jirsa},\ and\ \citenamefont {Kelso}}]{assisi2005synchrony}%
  \BibitemOpen
  \bibfield  {author} {\bibinfo {author} {\bibfnamefont {C.~G.}\ \bibnamefont
  {Assisi}}, \bibinfo {author} {\bibfnamefont {V.~K.}\ \bibnamefont {Jirsa}}, \
  and\ \bibinfo {author} {\bibfnamefont {J.~S.}\ \bibnamefont {Kelso}},\
  }\href@noop {} {\bibfield  {journal} {\bibinfo  {journal} {Physical review
  letters}\ }\textbf {\bibinfo {volume} {94}},\ \bibinfo {pages} {018106}
  (\bibinfo {year} {2005})}\BibitemShut {NoStop}%
\bibitem [{\citenamefont {Lafuerza}\ \emph {et~al.}(2010)\citenamefont
  {Lafuerza}, \citenamefont {Colet},\ and\ \citenamefont
  {Toral}}]{lafuerza2010nonuniversal}%
  \BibitemOpen
  \bibfield  {author} {\bibinfo {author} {\bibfnamefont {L.~F.}\ \bibnamefont
  {Lafuerza}}, \bibinfo {author} {\bibfnamefont {P.}~\bibnamefont {Colet}}, \
  and\ \bibinfo {author} {\bibfnamefont {R.}~\bibnamefont {Toral}},\
  }\href@noop {} {\bibfield  {journal} {\bibinfo  {journal} {Physical review
  letters}\ }\textbf {\bibinfo {volume} {105}},\ \bibinfo {pages} {084101}
  (\bibinfo {year} {2010})}\BibitemShut {NoStop}%
\bibitem [{\citenamefont {Mejias}\ and\ \citenamefont
  {Longtin}(2014)}]{mejias2014differential}%
  \BibitemOpen
  \bibfield  {author} {\bibinfo {author} {\bibfnamefont {J.~F.}\ \bibnamefont
  {Mejias}}\ and\ \bibinfo {author} {\bibfnamefont {A.}~\bibnamefont
  {Longtin}},\ }\href@noop {} {\bibfield  {journal} {\bibinfo  {journal}
  {Frontiers in computational neuroscience}\ }\textbf {\bibinfo {volume} {8}},\
  \bibinfo {pages} {107} (\bibinfo {year} {2014})}\BibitemShut {NoStop}%
\bibitem [{\citenamefont {Tripathy}\ \emph {et~al.}(2013)\citenamefont
  {Tripathy}, \citenamefont {Padmanabhan}, \citenamefont {Gerkin},\ and\
  \citenamefont {Urban}}]{tripathy2013intermediate}%
  \BibitemOpen
  \bibfield  {author} {\bibinfo {author} {\bibfnamefont {S.~J.}\ \bibnamefont
  {Tripathy}}, \bibinfo {author} {\bibfnamefont {K.}~\bibnamefont
  {Padmanabhan}}, \bibinfo {author} {\bibfnamefont {R.~C.}\ \bibnamefont
  {Gerkin}}, \ and\ \bibinfo {author} {\bibfnamefont {N.~N.}\ \bibnamefont
  {Urban}},\ }\href@noop {} {\bibfield  {journal} {\bibinfo  {journal}
  {Proceedings of the National Academy of Sciences}\ }\textbf {\bibinfo
  {volume} {110}},\ \bibinfo {pages} {8248} (\bibinfo {year}
  {2013})}\BibitemShut {NoStop}%
\bibitem [{\citenamefont {Beiran}\ \emph {et~al.}(2018)\citenamefont {Beiran},
  \citenamefont {Kruscha}, \citenamefont {Benda},\ and\ \citenamefont
  {Lindner}}]{beiran2018coding}%
  \BibitemOpen
  \bibfield  {author} {\bibinfo {author} {\bibfnamefont {M.}~\bibnamefont
  {Beiran}}, \bibinfo {author} {\bibfnamefont {A.}~\bibnamefont {Kruscha}},
  \bibinfo {author} {\bibfnamefont {J.}~\bibnamefont {Benda}}, \ and\ \bibinfo
  {author} {\bibfnamefont {B.}~\bibnamefont {Lindner}},\ }\href@noop {}
  {\bibfield  {journal} {\bibinfo  {journal} {Journal of computational
  neuroscience}\ }\textbf {\bibinfo {volume} {44}},\ \bibinfo {pages} {189}
  (\bibinfo {year} {2018})}\BibitemShut {NoStop}%
\bibitem [{\citenamefont {Padmanabhan}\ and\ \citenamefont
  {Urban}(2010)}]{padmanabhan2010intrinsic}%
  \BibitemOpen
  \bibfield  {author} {\bibinfo {author} {\bibfnamefont {K.}~\bibnamefont
  {Padmanabhan}}\ and\ \bibinfo {author} {\bibfnamefont {N.~N.}\ \bibnamefont
  {Urban}},\ }\href@noop {} {\bibfield  {journal} {\bibinfo  {journal} {Nature
  neuroscience}\ }\textbf {\bibinfo {volume} {13}},\ \bibinfo {pages} {1276}
  (\bibinfo {year} {2010})}\BibitemShut {NoStop}%
\bibitem [{\citenamefont {Shamir}\ and\ \citenamefont
  {Sompolinsky}(2006)}]{shamir2006implications}%
  \BibitemOpen
  \bibfield  {author} {\bibinfo {author} {\bibfnamefont {M.}~\bibnamefont
  {Shamir}}\ and\ \bibinfo {author} {\bibfnamefont {H.}~\bibnamefont
  {Sompolinsky}},\ }\href@noop {} {\bibfield  {journal} {\bibinfo  {journal}
  {Neural computation}\ }\textbf {\bibinfo {volume} {18}},\ \bibinfo {pages}
  {1951} (\bibinfo {year} {2006})}\BibitemShut {NoStop}%
\bibitem [{\citenamefont {Pfeil}\ \emph {et~al.}(2016)\citenamefont {Pfeil},
  \citenamefont {Jordan}, \citenamefont {Tetzlaff}, \citenamefont {Gr{\"u}bl},
  \citenamefont {Schemmel}, \citenamefont {Diesmann},\ and\ \citenamefont
  {Meier}}]{pfeil2016effect}%
  \BibitemOpen
  \bibfield  {author} {\bibinfo {author} {\bibfnamefont {T.}~\bibnamefont
  {Pfeil}}, \bibinfo {author} {\bibfnamefont {J.}~\bibnamefont {Jordan}},
  \bibinfo {author} {\bibfnamefont {T.}~\bibnamefont {Tetzlaff}}, \bibinfo
  {author} {\bibfnamefont {A.}~\bibnamefont {Gr{\"u}bl}}, \bibinfo {author}
  {\bibfnamefont {J.}~\bibnamefont {Schemmel}}, \bibinfo {author}
  {\bibfnamefont {M.}~\bibnamefont {Diesmann}}, \ and\ \bibinfo {author}
  {\bibfnamefont {K.}~\bibnamefont {Meier}},\ }\href@noop {} {\bibfield
  {journal} {\bibinfo  {journal} {Physical Review X}\ }\textbf {\bibinfo
  {volume} {6}},\ \bibinfo {pages} {021023} (\bibinfo {year}
  {2016})}\BibitemShut {NoStop}%
\bibitem [{\citenamefont {Horsthemke}\ and\ \citenamefont
  {Lefever}(2006)}]{Lefever2006}%
  \BibitemOpen
  \bibfield  {author} {\bibinfo {author} {\bibfnamefont {W.}~\bibnamefont
  {Horsthemke}}\ and\ \bibinfo {author} {\bibfnamefont {R.}~\bibnamefont
  {Lefever}},\ }\href@noop {} {\emph {\bibinfo {title} {{Noise-Induced
  Transitions (2nd edition)}}}}\ (\bibinfo  {publisher} {Springer-Verlag, New
  York},\ \bibinfo {year} {2006})\BibitemShut {NoStop}%
\bibitem [{\citenamefont {Lindner}\ \emph {et~al.}(2004)\citenamefont
  {Lindner}, \citenamefont {Garc{\i}a-Ojalvo}, \citenamefont {Neiman},\ and\
  \citenamefont {Schimansky-Geier}}]{lindner2004effects}%
  \BibitemOpen
  \bibfield  {author} {\bibinfo {author} {\bibfnamefont {B.}~\bibnamefont
  {Lindner}}, \bibinfo {author} {\bibfnamefont {J.}~\bibnamefont
  {Garc{\i}a-Ojalvo}}, \bibinfo {author} {\bibfnamefont {A.}~\bibnamefont
  {Neiman}}, \ and\ \bibinfo {author} {\bibfnamefont {L.}~\bibnamefont
  {Schimansky-Geier}},\ }\href@noop {} {\bibfield  {journal} {\bibinfo
  {journal} {Physics reports}\ }\textbf {\bibinfo {volume} {392}},\ \bibinfo
  {pages} {321} (\bibinfo {year} {2004})}\BibitemShut {NoStop}%
\end{thebibliography}

%

\end{document}